# Dynamic Equivalent Identification of Interconnected Areas Using Disturbance records from Synchro phasor Measurement Units

# Identificación de Equivalentes Dinámicos de Áreas Interconectadas empleando Registros de Perturbaciones de Unidades de Medición Sincrofasorial

W.F. Brito[1]   0009-0004-3002-2427   M.S. Chamba[1,2]   0000-0001-6843-7151
W Vargas[1]    0000-0002-4239-2511   D.E. Echeverria[1,2]   0000-0002-1743-9234

[1]Pontificia Universidad Católica del Ecuador - Sede Esmeraldas, Esmeraldas, Ecuador
E-mail: wfbrito@pucese.edu.ec, mschamba@puecese.edu.ec, deecheverriaj@pucese.edu.ec, wavargas@pucese.edu.ec
[2]Operador Nacional de Electricidad - CENACE, Quito, Ecuador
E-mail: mchamba@cenace.gob.ec. decheverria@cenace.gob.ec

*Abstract*

In this paper, a methodology for modelling a dynamic equivalent of an external area is presented. The equivalent consists of a generator with series impedance and a parallel load (generalized Ward equivalent), integrating control systems such as the Automatic Voltage Regulator (AVR) and the Speed Regulator (GOV) in a test system known as PST-16. The main objective is to identify the parameters of the control systems and other parameters inherent to the generator so that the response of the equivalent system is similar to the response of the complete system.

*Resumen*

En este documento, se presenta una metodología para modelar un equivalente dinámico de un área externa. El equivalente consta de un generador con impedancia serie y una carga en paralelo (equivalente Ward generalizado), integrando sistemas de control como el Regulador Automático de Voltaje (AVR) y el Regulador de Velocidad (GOV) en un sistema de prueba conocido como PST-16. El objetivo principal es identificar los parámetros de los sistemas de control y otros parámetros inherentes al generador para que la respuesta del sistema equivalente sea similar a la respuesta del sistema completo.

*Palabras clave—* Equivalente, Externo, Dinámico, Ward.

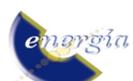

# 1. INTRODUCCIÓN

La planificación y el funcionamiento de un Sistema Eléctrico de Potencia (SEP) se fundamentan en una serie de evaluaciones que abordan simulaciones tanto en el funcionamiento en estado estacionario como en el dinámico. En la actualidad, los SEPs incorporan una amplia gama de componentes, producto de la expansión del propio sistema como de la creciente interconexión de los sistemas, que deben ser modelados de manera precisa para reflejar el comportamiento real del SEP. La abundancia de estos componentes conduce a un aumento significativo en la cantidad de ecuaciones no lineales algebraicas diferenciales, lo que significa que el problema que debe resolverse se vuelve más complicado demandando una mayor capacidad de cómputo y mayores tiempos de simulación [1].

Con al afán de reducir los tiempos de simulación y comprensión de los fenómenos dinámicos se han planteado varios trabajos que usan equivalentes de red de una región del sistema, las cuales deben reproducir el mismo comportamiento estacionario y dinámico. En [2], [3], [4], [5], [6], [7] y [8] se han desarrollado múltiples métodos para encontrar equivalentes de red en los que se divide a un sistema muy grande en dos o más áreas, siendo el área de interés para estudios denominada como área interna y el resto de las áreas como externas.

En [2] se propone un equivalente de red denominado equivalente Ward. En este equivalente se reemplaza el área externa por una admitancia fija equivalente en las barras de enlace entre las dos áreas. El flujo de potencia entre las áreas con esta nueva admitancia debe coincidir con el flujo de potencia inicial. En estudios de contingencias, este equivalente demuestra ser razonablemente bueno, sin embargo no logra representar de manera adecuada el comportamiento de las barras PV como se menciona en [3], pues no considera el comportamiento dinámico de las barras de generación.

En [4] se hace mención a un equivalente Ward en el cual se sugiere que no se reduzcan las barras de tipo PV del sistema externo y reducir solamente las barras de tipo PQ con la finalidad de tener una mejor representación de la inyección de reactivos de estas barras en los estudios del sistema. En esta metodología, si la red externa está compuesta por muchas barras del tipo PV entonces la reducción del sistema es baja y tampoco se consideran el comportamiento dinámico de las barras PV. Adicionalmente, en [4] se presenta un equivalente Ward generalizado, en el cual se reemplaza a la red externa por una barra PQ y PV interconectadas. En este trabajo, el generador de la barra PV es modelado junto con un AVR para representar de forma más precisa la inyección de reactivos de los generadores externos. En esta aproximación no se toman en cuenta los reguladores de velocidad y voltaje, por lo que esta aproximación no representa adecuadamente eventos dinámicos.

En [5] se extiende el equivalente Ward propuesto en [2] añadiendo barras PV ficticias. Estas barras, cuando las condiciones del sistema cambian responden absorbiendo o generando potencia. Este equivalente es preciso en cuanto a los cálculos de los flujos de reactivos; no obstante, en estudios dinámicos su precisión es reducida.

El Equivalente Radial Independiente (REI, por sus siglas en inglés) se propone en [6], en donde agregan las corrientes de las barras del sistema externo inyectándolas en una única barra. Luego, las corrientes inyectadas son distribuidas en estas barras a través de una red radial denominada red REI. Los estudios de flujos de potencia con este equivalente coinciden con gran precisión con el de la red no reducida pero para una condición diferente de operación del sistema, el equivalente REI obtenido debe ser recalculado como se indica en [3].

En [7] se propone representar a un sistema vecino como un equivalente dinámico compuesto por un generador, sus sistemas de control (AVR, GOV y PSS) y una carga en paralelo. En este trabajo se menciona que existen dos metodologías principales para encontrar este equivalente: i) la agregación teórica y ii) la identificación en línea. La agregación teórica requiere de un conocimiento profundo del sistema a reducir; mientras que, la identificación en línea puede llegar a necesitar de una gran capacidad de cómputo.

En [8] se emplea una metodología similar a la de [7], identificando los parámetros del generador equivalente; sin embargo, la metodología usada es fuera de línea, implementando la reproducción de eventos o "Play Back" por su nombre en inglés. Esta metodología facilita encontrar los equivalentes dinámicos cuando la información del área externa a representar es escasa.

Los sistemas eléctricos cada vez son más complejos y extensos, esto dificulta contar con datos suficientes para modelar el sistema sobre todo en áreas, cuyo control está a cargo de entidades diferentes (como lo pueden ser dos sistemas interconectados), muy lejanas o áreas en donde no se cuente con sistemas de medición efectivos. En este sentido, la presente investigación propone desarrollar una metodología para realizar la Identificación de Equivalentes Dinámicos de Áreas Interconectadas empleando simulaciones que simulen los Registros de Perturbaciones de Unidades de Medición Sincrofasorial (PMU).

Para ello, en la segunda sección de este documento, se propone la revisión de los principales conceptos para comprender la propuesta, tales como: optimización heurística, sistemas de control, sistemas de medición sicrofasorial, entre otros. En la tercera sección se presenta la metodología que abarca la identificación de los equivalentes en estado estacionario y dinámico. En la cuarta sección se presenta la discusión y análisis de resultados. Finalmente, se presentan las principales conclusiones y recomendaciones.

## 2. MARCO TEÓRICO

### 2.1 Optimización Heurística

Los algoritmos heurísticos son enfoques computacionales que se emplean para aproximarse a soluciones óptimas en problemas complejos. Estos métodos se sustentan en reglas prácticas y buscan iterativamente mejorar una solución mediante una medida de calidad. Es importante destacar que, estos algoritmos hacen un mínimo de suposiciones sobre el problema a optimizar y realizan una búsqueda en un espacio muestral grande, con un costo computacional razonable [9]. En el ámbito de la investigación, es común el uso de librerías establecidas para la implementación algoritmos heurísticos, entre las cuales Pymoo es una de las más destacadas.

Pymoo es un paquete de optimización multiobjetivo desarrollado en el lenguaje de programación Python, destacándose por su capacidad para abordar una amplia gama de problemas que involucran la optimización de múltiples criterios. Este paquete brinda acceso a una variedad de algoritmos de optimización mono y multiobjetivo, entre ellos, se incluyen métodos como: el Algoritmo de Optimización de Enjambre de Partículas (PSO), Diferencial Evolutivo (DE), NSGA-II (Non-Dominated Sorting Genetic Algorithm II), NSGA-III, SPEA2 (Strength Pareto Evolutionary Algorithm 2), y MOEA/D (Multi-Objective Evolutionary Algorithm Based on Decomposition), entre otros [10]. La versatilidad y potencia de Pymoo lo convierten en una herramienta esencial para encontrar soluciones óptimas [10].

### 2.2 Sistemas de Control del Generador

Los generadores eléctricos están equipados con sistemas de control de generación, que incluyen al AVR (Automatic Voltage Regulator) y GOV (Governor) como se indica en la Figura 1, los cuales desempeñan un papel esencial en la dinámica de producción de energía eléctrica. El AVR se encarga de mantener el voltaje dentro de límites aceptables y el GOV ajusta la velocidad y por consiguiente la frecuencia del generador según lo requiera el sistema. Estos sistemas trabajan en conjunto para garantizar un suministro eléctrico confiable y eficiente, contribuyendo así al funcionamiento óptimo de las redes eléctrica.

#### 2.2.1 Regulador Automático de Voltaje (AVR)

El Regulador Automático de Voltaje, AVR por sus siglas en inglés, es un dispositivo electrónico que se encarga de mantener en un valor constante el voltaje en los terminales del generador dentro de los valores permitidos. El AVR actúa sobre la excitatriz aumentando o disminuyendo la corriente de campo según sea requerido [11], [12]. Los modelos de AVRs estandarizados se encuentran disponibles en [13], estos modelos proporcionan una gran presión al momento de replicar el comportamiento de sistemas físicos.

#### 2.2.2 Regulador de Velocidad (GOV)

En un generador sincrónico, la ecuación que describe al movimiento está dada por (1)

$$2H\frac{d\omega_r}{dt} = T_m - T_e \quad (1)$$

Donde:

$H$: Constante de inercia (segundos)

$T_m$: Torque mecánico o Torque de la turbina (p.u)

$T_e$: Torque eléctrico o sincrónico (p.u)

$\omega_r$: Velocidad del rotor

Cuando existe un desbalance instantáneo entre la potencia eléctrica (*Pe*) y la potencia de carga (*PL*) se produce un cambio en el torque eléctrico de salida (*Te*) del generador, produciendo una desigualdad con el torque mecánico (*Tm*). Esta desigualdad de torques provoca una variación de la velocidad angular del rotor y, por ende, de la frecuencia.

El regulador de velocidad, GOV por sus siglas en inglés, es un dispositivo de control de los generadores que aumenta o disminuye la velocidad del generador. El GOV está constantemente midiendo la velocidad del generador y envía una señal de control al sistema de válvulas para regular el flujo de energía primaria (vapor si es generador térmico o agua si es hidroeléctrico) que llega a la turbina. Como consecuencia de esta acción el Torque mecánico (Tm) se iguala al Torque eléctrico (Te) [14], [15].

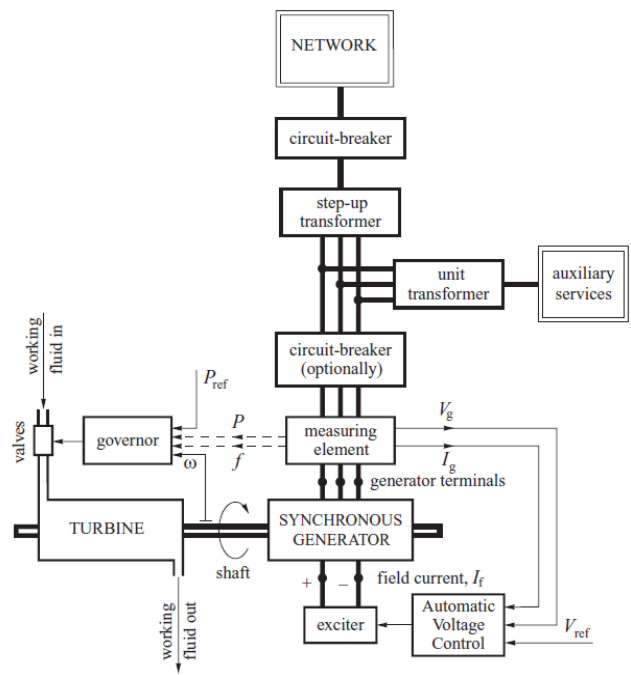

**Figura 1** Diagrama de Bloques de una Unidad de Generación [16]

Modelos estandarizados de reguladores de velocidad y turbinas estandarizados están disponibles en [17]. En este trabajo se optó por usar la turbina no lineal con columna de agua y un regulador de velocidad genérico tipo PID.

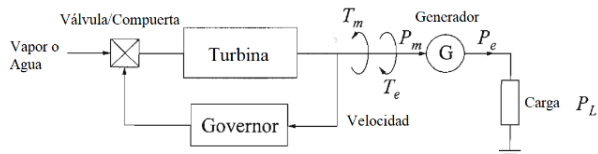

**Figura 2 Esquema de generación para regulación de frecuencia primaria [14]**

### 2.3 Optimización Multi Objetivo

En la identificación de parámetros para sistemas de control en equivalentes dinámicos, la optimización multiobjetivo desempeña un papel crucial al buscar soluciones que equilibren eficientemente diversos objetivos. La interacción entre eventos destaca la necesidad de este enfoque, ya que la respuesta de un sistema de control ante un evento puede tener implicaciones en su respuesta ante otros eventos. Este trabajo de investigación se centra en la implementación de la suma ponderada de pesos como método específico para la optimización multiobjetivo.

#### 2.3.1 Suma Ponderada de Pesos

El método de la suma ponderada de pesos combina a todas las funciones objetivo dentro de una sola función que está dada por la ecuación (2):

$$F(x) = w_1 f_1(x) + w_2 f_2(x) + \cdots + w_n f_n(x) \quad (2)$$

Donde:

$f_n$: Funciones objetivo. Ecuaciones que relacionan resultados para minimizar los errores de estado estacionario y dinámico

$w_n$: Coeficientes de peso

La importancia de cada función objetivo es ajustada mediante la multiplicación de su respectivo peso según lo requieran los requisitos específicos de cada problema. Los factores de peso deben ser positivos y satisfacer la expresión (3):

$$\sum_{i=1}^{M} w_i = 1, \quad w_i \in (0,1) \quad (3)$$

#### 2.3.2 Índices de Rendimiento (Funciones Objetivo)

Para lograr el objetivo de minimizar el error entre la señal real (mediciones PMUs) y la señal simulada existen múltiples criterios. La respuesta de un sistema de control se considera óptima cuando el ajuste de sus parámetros resulta en la minimización de su índice de rendimiento que compara la señal simulada con la señal deseada que en este caso puede ser la señal medida. Los distintos índices de desempeño se presentan en la Tabla 1.

Estos criterios permiten convertir al problema de sintonización en uno de optimización, en donde: $e(t)$ es la diferencia entre la señal obtenida (mediante simulación o medición) y una señal ideal de tipo escalón unitario o rampa unitaria, cuyo valor debe ser minimizado [18], [19].

**Tabla 1 Índices de desempeño**

| Índice de Desempeño | Descripción | Expresión |
|---|---|---|
| ISE | Criterio de la integral del error al cuadrado | $\int_0^t e^2(t)\,dt$ |
| ITSE | Integral del error cuadrado multiplicado por el tiempo | $\int_0^t te^2(t)\,dt$ |
| IAE | Criterio de la integral del valor absoluto del error | $\int_0^t \lvert e(t) \rvert\,dt$ |
| ITAE | Criterio de la integral del valor absoluto del error multiplicado por el tiempo | $\int_0^t t\lvert e(t) \rvert\,dt$ |

El criterio ISE es estricto con los grandes errores iniciales lo que lo hace ideal para identificar señales con sobreoscilaciones (overshoots), utilizado en las funciones objetivo del presente trabajo.

El criterio ITSE tiene errores iniciales grandes debido a su dependencia del tiempo. Este criterio es más estricto a medida que avanza el tiempo ya que la penalización es más severa. El criterio IAE es fácil de implementar, no obstante, su desempeño no es bueno, respecto al resto de índices al considerarse un índice demasiado básico.

Por último, el criterio ITAE tiene la particularidad de ser permisivo con los grandes errores iniciales e incrementa su severidad con los errores que ocurren más adelante en la señal. Estas características del criterio ITAE permiten una mejor selectividad en la búsqueda de los parámetros del sistema como se menciona en [20], [21], [22], lo que lo convierte en uno de los más recomendables para usarse en este tipo de problemas de optimización.

### 2.4 Sistemas de Medición de Áreas Extendida (WAMS)

Los Sistemas de medición de área extensa, en inglés Wide Area Measurement Systems, son sistemas avanzados para el monitoreo de variables de campo (voltaje y corriente) en los sistemas eléctricos de potencia. Además, estos sistemas cuentan con equipos de sistemas de posicionamiento global GPS para sincronizar el tiempo de todas las mediciones tomadas en los distintos puntos del WAMS. En la Figura 3 se presenta un esquema de la arquitectura del WAMS, donde el principal componente es la Unidad de Medición Fasorial (PMU).

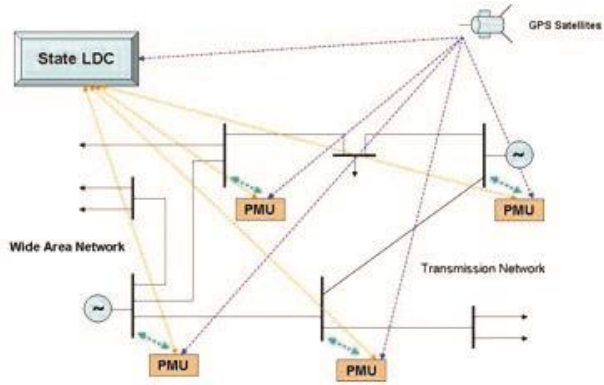

**Figura 3 Arquitectura WAMS [23]**

La Unidad de Medición Fasorial, PMU por sus siglas en inglés, es un dispositivo de alta precisión usado no solo en la medición de magnitud de señales sinusoidales de corriente o voltaje, sino que además permite la estimación de los fasores de dichas señales. Esta estimación se logra mediante la sincronización temporal con una señal GPS, asegurando que todas las mediciones se registren de manera simultánea y permitiendo así una comparación y una estimación precisas como se indica en la Figura 4. Además, las PMUs desempeñan un papel fundamental al transmitir estos datos al servidor del WAMS para supervisión y análisis por parte de los operadores del sistema [24].

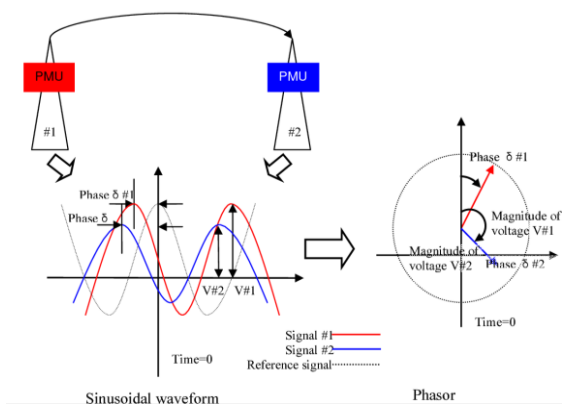

**Figura 4 Señales recibidas por las PMU [25]**

Debido a que estos sistemas están ampliamente implementados en las redes eléctricas, la metodología que se plantea en este trabajo se basará en mediciones que emulen el registro de una PMU para que, posteriormente la metodología pueda ser validada con mediciones de eventos dinámicos reales.

### 2.5 Sistema de Prueba

El sistema de prueba dinámico PST-16, tal como se muestra en la Figura 5, está compuesto por una red de 400/220 kV con 16 generadores [26] con sus respectivos sistemas de control (GOV, AVR, PSS), distribuidos en tres áreas interconectadas (A, B y C) mediante líneas de transmisión largas. Este estudio se enfoca en la sustitución de dos áreas, A y C, a través de generadores equivalentes como se ilustra en la Figura 6.

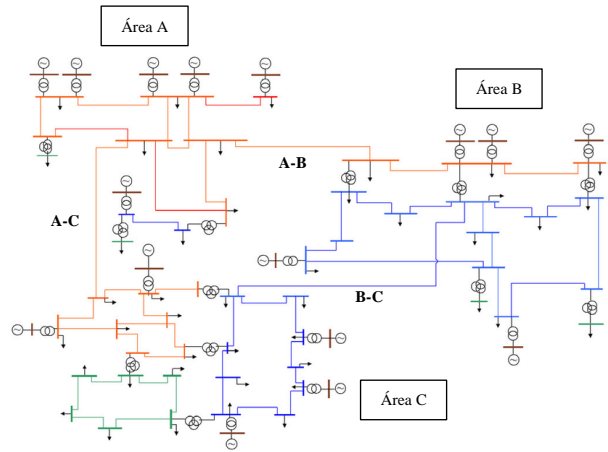

**Figura 5 Diagrama unifilar del Sistema PST-16 [26]**

Al ser dos áreas enlazadas (A y B) serán reemplazadas por un equivalente dinámico; para ello, debe existir una impedancia común entre los equivalentes Ward que reemplazaran a dichas áreas como se menciona en [27]. Tomando en cuenta este aspecto el equivalente de red se presenta en la Figura 6.

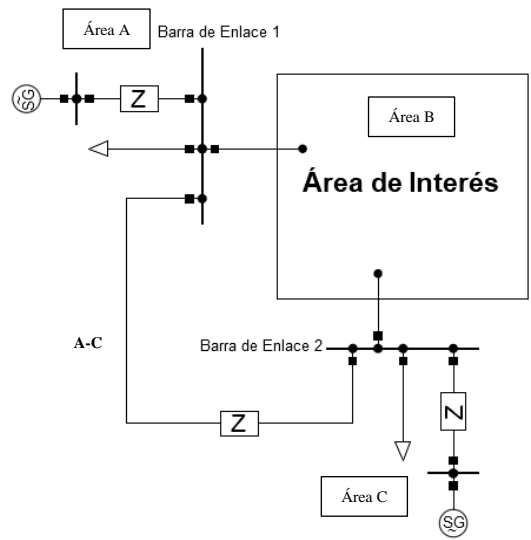

**Figura 6 Diagrama del Equivalente Dinámico**

### 3. METODOLOGÍA PROPUESTA

Con la finalidad de simplificar el modelado del equivalente dinámico de la red, la metodología propuesta asume los siguientes criterios:

- Los generadores equivalentes se modelan según el enfoque del "Modelo Ward generalizado". Esto implica la representación del generador como una combinación de un generador con impedancia equivalente en serie y una carga en paralelo.
- Los sistemas de regulación de voltaje y velocidad se escogen considerando que, los mismos tengan un enlace de retroalimentación. Esta elección permite capturar de manera efectiva las dinámicas asociadas con la regulación en el sistema.
- La carga eléctrica conectada en paralelo se modela como un elemento de potencia constante.
- Se desprecian las resistencias y reactancias internas de los generadores equivalentes, siendo estas asumidas por la impedancia serie equivalente al igual que los valores de característicos de las líneas de transmisión, mientras que los demás parámetros se mantienen con sus valores por defecto.
- La potencia de corto circuito en las barras de enlace debe ser similar al del sistema completo.
- Al implementar el equivalente dinámico, los flujos por las líneas conectadas a la barra de enlace entre las áreas deben ser lo más similar posible al obtenido previo a la implementación del equivalente.
- Se consideran dos eventos dentro del área de interés, un incremento en la carga general 18 de un 30 % y un evento de corto circuito con desconexión de línea en 100 ms en la línea B3b-B6 al 50% de la línea.

La identificación del equivalente dinámico se realiza en dos etapas, en la primera se realiza una identificación en estado estacionario, es decir se identifican los parámetros para que el flujo de potencia con el equivalente sea el mismo que con el sistema completo y así mismo, que la potencia de corto circuito en las barras de enlace sean los mismos que con el sistema completo.

En la segunda etapa se ajusta solamente los parámetros dinámicos del equivalente, es decir la constate de inercia y parámetros de los reguladores de voltaje y velocidad.

### 3.1 Identificación en Estado Estacionario

En esta etapa, se lleva a cabo la identificación de parámetros de las impedancias serie y común en los escenarios a través de cálculos de flujo de potencia y cortocircuito. Asimismo, se determinan las magnitudes de las potencias nominales correspondientes a los generadores equivalentes y las cargas en paralelo. Esta tarea implica la creación de una función multiobjetivo, como se muestra en (4), para minimizar los errores entre los flujos de potencia y cortocircuitos. Las funciones objetivo utilizadas consideran el criterio ISE, para minimizar el error cuadrático medio de la medición del flujo de potencia y cortocircuitos de la barra de enlace.

$$F_1(x) = w_1 F_{PF}(x) + w_2 F_{SHC}(x) \quad (4)$$

Donde:

$w_1, w_2$: Coeficientes de peso.

$F_{PF}(x), F_{SHC}(x)$: Funciones Objetivo. Ecuaciones que relacionan los resultados de flujos de potencia y cortocircuitos, respectivamente.

El procedimiento de esta identificación es resumido en la Figura 7, comenzando con la aplicación de parámetros de impedancias y potencias, seguido de un flujo de potencia con el equivalente de red. Luego, se comparan los flujos de potencia de los elementos adyacentes en las barras de enlace (con red completa y con el equivalente) para definir la primera función objetivo. Después, se realiza un cortocircuito y se comparan los aportes de los elementos adyacentes de la misma manera que en el flujo de potencia. La función objetivo global se define combinando ambas funciones objetivo, y se procede a minimizarla hasta cumplir el criterio de parada.

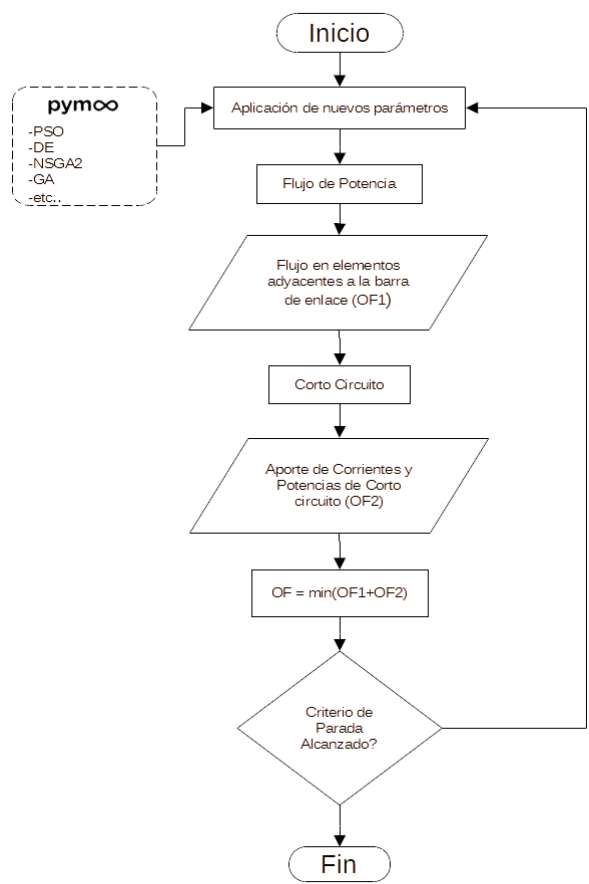

**Figura 7** Diagrama de Flujo de la Identificación en estado estacionario

La inyección de potencia (activa o reactiva) del generador equivalente, impedancia común y carga debe igualar la inyección de potencia de las líneas reemplazadas en las barras de enlace y mantener constante el flujo de potencia en las líneas y elementos no reemplazados. Simultáneamente, la inyección de potencia de cortocircuito del equivalente debe igualar la inyección con respecto al sistema completo. Las funciones objetivo en ambos escenarios buscan minimizar el error en los flujos y cortocircuitos entre el sistema original y su equivalente. Los pesos asignados a cada función objetivo son iguales a 0.5, dando a entender que son igual de importantes ambos objetivos.

### 3.2 Identificación en Estado Dinámico

Durante esta etapa, se simulan eventos de frecuencia (incremento de carga) y voltaje (cortocircuito con despeje de línea). Las señales de voltaje, potencia y frecuencia en las líneas conectadas a las barras de enlace y con el sistema completo se almacenan simulando las mediciones de las unidades de medición sincrofasorial (PMU). Estas señales se emplean posteriormente como referencia para identificar los parámetros de inercia del generador y de los sistemas de control de velocidad y voltaje, definiéndose así la función objetivo que se presenta en (5). Las funciones objetivo utilizadas consideran el criterio ISE, debido a que penalizan de manera equitativa a todos los puntos de una referencia dinámica.

$$F_2(x) = w_1 f_{Frec}(x) + w_2 f_{volt}(x) \quad (5)$$

Donde:

$w_1, w_2$: Coeficientes de peso.

$f_{Frec}(x), f_{volt}(x)$: Ecuaciones que relacionan los resultados de eventos de frecuencia y voltaje, respectivamente.

Las funciones objetivo-individuales pretenden reducir el error entre la señal de respuesta del sistema equivalente y la señal de referencia. Para este trabajo se prioriza la respuesta ante eventos de frecuencia del sistema, asignando pesos de 0.8 para eventos de frecuencia y 0.2 para eventos de voltaje.

El proceso de identificación dinámico se resume en la Figura 8. Inicialmente, se eligen los eventos (frecuencia y voltaje) a replicar. Luego, se establece la conexión entre Power Factory y Python, utilizando la librería Pymoo para seleccionar un algoritmo heurístico. Se define el índice de rendimiento como función objetivo y se aplican los nuevos parámetros para llevar a cabo una simulación donde se comparan los nuevos eventos simulados con la red equivalente y las referencias obtenidas de las simulaciones con el sistema completo. Una vez que se alcanza el criterio de parada, se considera identificados los parámetros óptimos y finaliza el algoritmo.

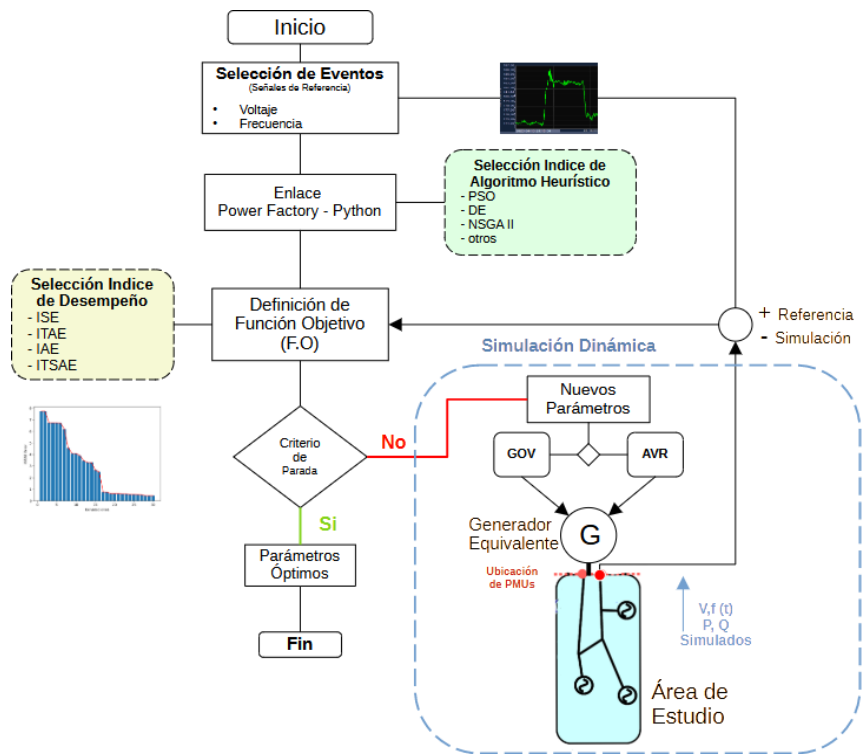

**Figura 8 Diagrama de Flujo de la Identificación Dinámica**

Para este trabajo se ha optado por usar el modelo de AVR IEEE AC5A debido a su simplicidad y presencia de lazo de retro alimentación, una turbina no lineal con columna de agua y un regulador de velocidad genérico tipo PID. Los diagramas de estos modelos están disponibles en las referencias [13] y [17], respectivamente.

### 4. DISCUSION Y RESULTADOS

En esta sección se presentan los resultados obtenidos de la identificación paramétrica.

### 4.1 Identificación en Estado Estacionario

Los flujos de potencia sobre los elementos del sistema se presentan en la Tabla 2, en esta se realiza una comparación de los flujos originales por los elementos en comparación con los flujos al implementar el equivalente. Es necesario recalcar que para las líneas A-B y B-C, las nuevas potencias provienen de la suma de los aportes por los elementos equivalentes (Generador, impedancias y cargas).

**Tabla 2 Flujos de Potencia**

| Elemento | Po | P_new | ΔP | Qo | Q_new | ΔQ |
|---|---|---|---|---|---|---|
| B1-B2 | -169.5 | -156.61 | 12.89 | -173.2 | -172.25 | 0.95 |
| B1 3T | 308 | 325.42 | 17.42 | -76.7 | -79.28 | 2.58 |
| A-B | -938.5 | -939.75 | 1.25 | 209.9 | 243.91 | 34.01 |
| B2c-B4 | 220.4 | 215.21 | 5.19 | 90.6 | 88.11 | 2.49 |
| B2 3T | -1569.8 | -1641.83 | 72.03 | -103.7 | -106.36 | 2.66 |
| B-C | 441.8 | 441.19 | 0.61 | -79.1 | -79.73 | 0.63 |
| B2c-B5 | 245 | 280.69 | 35.69 | -9.2 | -10.44 | 1.24 |
| B2c-B6 | 89.6 | 84.34 | 5.26 | 41.4 | 44.68 | 3.28 |

En la Tabla 3 se presentan las potencias y corrientes de corto circuito antes y después de implementarse el sistema equivalente.

**Tabla 3 Cortocircuito en la Barra de enlace**

| Elemento | Skss | Skss_new | ΔSkss | Ikss | Ikss_new | ΔIkss |
|---|---|---|---|---|---|---|
| B1 | 8028.1 | 7526 | 502.1 | 12.197 | 11.435 | 0.762 |
| B2c | 13879.4 | 12993.8 | 885.6 | 36.424 | 34.1 | 2.324 |

En la Tabla 4 se presentan los valores de las impedancias se secuencia positiva y negativa (que para este trabajo se consideran iguales) y la de secuencia cero. En la Tabla 5 se presentan en cambio los valores de las conductancias y susceptancias de las impedancias equivalentes.

**Tabla 4 Impedancias de secuencia equivalentes**

| Z | S Base (MVA) | Sec. (+/-) | | Sec. 0 | |
|---|---|---|---|---|---|
| | | Real | Imaginaria | Real | Imaginaria |
| Z Eq 1 | 5000 | 0.5130134 | 3.303684 | 1.311948 | 0.4900606 |
| Z Eq 2 | 5000 | 0.00717906 | 3.495438 | 2.862936 | 1.47457 |
| Z 1-2 | 1000 | 0.8268194 | 3.315204 | 0.7496988 | 0.4584266 |

**Tabla 5 Conductancia y Suceptancia Equivalentes**

| | S Base (MVA) | Sec. (+/-) | | Sec. (+/-) | |
|---|---|---|---|---|---|
| | | Conduct. i | Sucept. i i | Conduct. j | Sucept. j |
| Z Eq 1 | 5000 | 0.00002507 | 0.01232465 | 0.00043594 | 0.00012577 |
| Z Eq 2 | 5000 | 0.00057239 | 0.01658769 | 0.00001463 | 0.01650044 |
| Z 1-2 | 1000 | 0.00041881 | 0.00000001 | 0.00014285 | 0.01701847 |

La potencia nominal del cada generador junto con los valores de flujo de potencia de cada carga conectada en paralelo se presenta en la Tabla 6.

**Tabla 6 Generación y Carga Equivalentes**

| | S Nom (MVA) | P (MW) | Q (MVAr) |
|---|---|---|---|
| G1 | 25153.4 | 0 | 0 |
| G2 | 29361.17 | 0 | 0 |
| Eq Load 1 | - | -944.8145 | 260.8259 |
| Eq Load 2 | - | 443.2095 | -24.47694 |

### 4.2 Identificación en Estado Dinámico

Los parámetros de los reguladores de voltaje de cada generador equivalente se presentan en la Tabla 7, mientras que los parámetros de los reguladores de velocidad y turbina en la Tabla 8

**Tabla 7 Parámetros de los AVRs Equivalentes**

| Parámetro | G. Eq 1 | G. Eq 2 | Unidad |
|---|---|---|---|
| Ka | 2.82901 | 1.00903 | [p.u.] |
| Ta | 0.3306341 | 0.1153288 | [s] |
| Te | 0.1516396 | 0.5137304 | [s] |
| Ke | 0.4065861 | 0.4707643 | [p.u.] |
| Efd1 | 3.778565 | 4.043252 | [p.u.] |
| SeEfd1 | 0.2829464 | 0.1258103 | [p.u.] |
| Efd2 | 3.733107 | 3.114192 | [p.u.] |
| SeEfd2 | 0.0145818 | 0.1186335 | [p.u.] |
| Kf | 0.01656878 | 0.1899858 | [p.u.] |
| Tf1 | 1.034084 | 1.314748 | [s] |
| Tf2 | 0.01547329 | 0.1354222 | [s] |
| Tf3 | 0.286422 | 0.5903848 | [s] |

**Tabla 8 Parámetros de los Reguladores de Velocidad y Turbina Equivalentes**

| Parámetro | G. Eq 1 | G. Eq 2 | Unidad |
|---|---|---|---|
| h | 1.522318 | 2.561282 | [s] |
| Rp | 0.03686194 | 0.00574989 | [p.u.] |
| Tt | 0.5509713 | 0.02583656 | [s] |
| db1 | 0 | 0.00002308 | [p.u.] |
| Td | 0.05091809 | 0.00276573 | [s] |
| Kp | 2.65504 | 5.952962 | [p.u.] |
| Ki | 0.05951383 | 0.03044037 | [p.u.] |
| Kd | 0.01139807 | 0.00987222 | [p.u.] |
| Tf | 0.1576367 | 0.07094841 | [s] |
| Tg | 0.02035356 | 0.08723536 | [s] |
| Tp | 0.01956299 | 0.2254878 | [s] |
| Tw | 0.485097 | 0.4214861 | [s] |

Una figura comparativa entre respuestas ante un evento de frecuencia con el sistema completo y la obtenida con el sistema equivalente se presentan en las Figuras 9 Y 10.

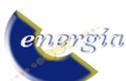

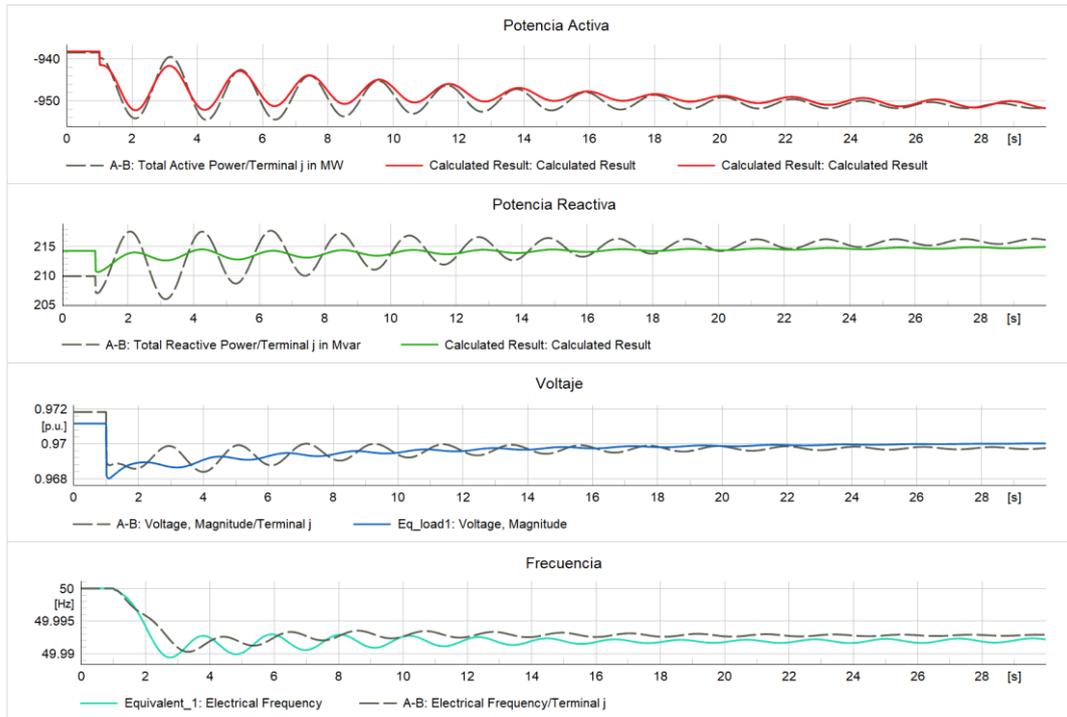

**Figura 9** Comparación de respuesta del sistema completo (gris entrecortado) v.s. Equivalente, Potencia activa (rojo), Potencia reactiva (verde), Voltaje (azul) y Frecuencia en la barra de enlace B1 (celeste)

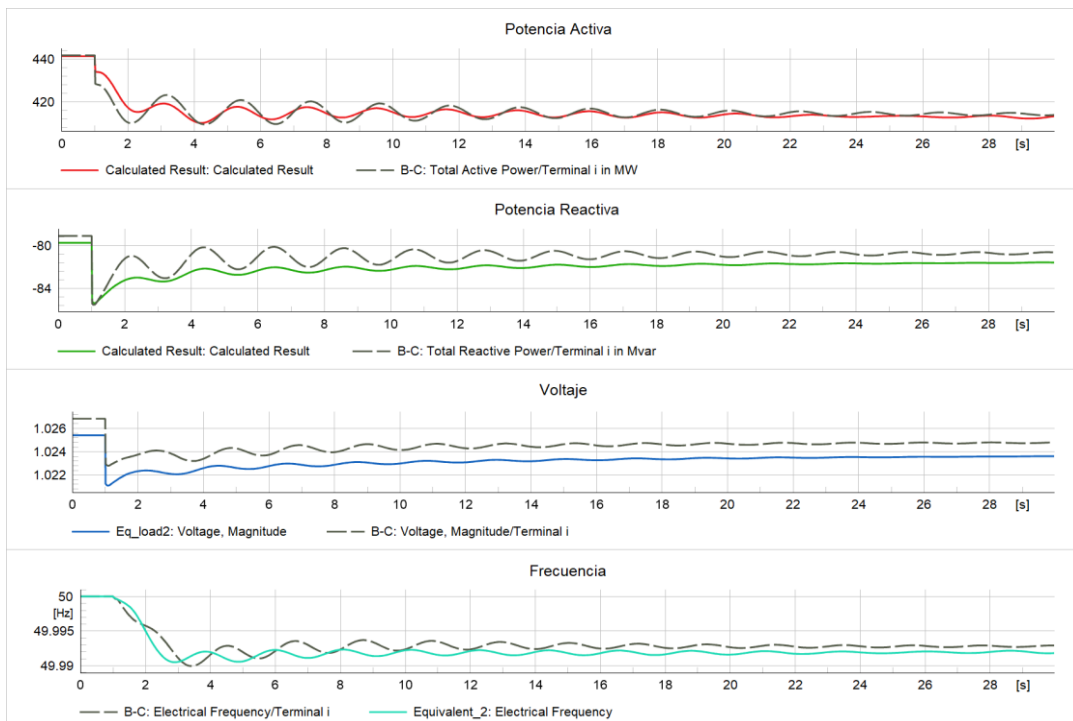

**Figura 10** Comparación de respuesta del sistema completo (gris entrecortado) v.s. Equivalente, Potencia activa (rojo), Potencia reactiva (verde), Voltaje (azul) y Frecuencia en la barra de enlace B2 (celeste)

Como se puede observar, la respuesta del evento de frecuencia tiene un menor error en comparación con la respuesta al evento de voltaje; esto debido a que, inicialmente se le asigno un mayor peso al evento de frecuencia por lo que su penalización fue más rigurosa al encontrar los parámetros óptimos.

Los valores elevados de la potencia base de los generadores equivalentes pueden atribuirse a las altas magnitudes de las impedancias equivalentes. La magnitud de estos valores se deriva de la amplitud de los límites de búsqueda establecidos por el algoritmo utilizado para identificar dichos parámetros.

## 5. CONCLUSIONES, RECOMENDACIONES Y FUTUROS TRABAJOS

El presente trabajo propone una novedosa metodología y desarrolla una herramienta robusta que permite realizar, tanto la identificación paramétrica en estado estacionario como en estado dinámico, de sistemas o áreas eléctricas. En esta metodología permite identificar dos puntos de interconexión de dos sistemas eléctricos, mediante impedancias, cargas y sistemas de generación con sistemas de control equivalentes. Para ello, se plantean modelos de optimización que permiten comparar las variables reales del sistema completo con las variables simuladas del sistema equivalente, mediante programación de una herramienta en Python que enlazando la simulación estacionarias y dinámicas con PowerFactory de DIgSILENT. Adicionalmente, es importante destacar que la herramienta es versátil puesto que, permite escoger diferentes algoritmos de optimización y diferentes funciones objetivo (índices de rendimiento), logrando de ampliar los espacios de búsqueda y comparar dichas respuestas.

En este trabajo se utilizó modelos simplificados de los sistemas de control para representar el comportamiento dinámico de las áreas reemplazadas por un equivalente. Debido a la simplicidad de los sistemas de control es posible que algunos fenómenos no sean representados de la forma mas adecuada, por lo que, es necesario utilziar modelos más complejos, así como una mayor cantidad de eventos para la identificación de los equivalentes dinámicos.

Reducir el error entre los dos tipos de eventos, voltaje y frecuencia es fundamental para una identificación adecuada, en este trabajo se priorizó la respuesta ante eventos de frecuencia para disminuir la complejidad del problema y reducir el costo computacional. Con este antecedente, se concluye que es necesario igualar o utilizar diferentes pesos de las dos funciones objetivo, así como reducir los espacios de búsqueda gradualmente para lograr una identificación más precisa.

En este trabajo solamente se consideró la identificación de los sistemas GOV y AVR, sin embargo, se modelo Estabilizadores de Sistemas Potencia (PSS, por sus siglas en inglés) con parámetros por defecto de librería. Este control actúa estabilizando el sistema en rangos de frecuencia específicos, permitiendo incrementar la precisión de la respuesta de los equivalentes dinámicos. Como futuro trabajo, se plantea definir una metodología que incluya la identificación de este sistema de control.

## 6. REFERENCIAS BIBLIOGRÁFICAS

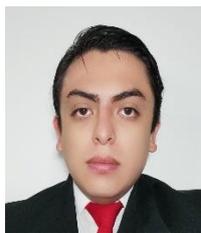

**Wilson Brito.**- Nació en Quito, Ecuador en 1995. Obtuvo el título de Ingeniero Eléctrico en la Escuela Politécnica Nacional, Ecuador en el 2022. Actualmente está realizando sus estudios de maestría en ingeniería eléctrica en la Universidad Católica del Ecuador.

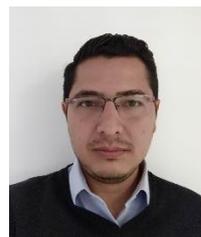

**Marlon Santiago Chamba.**- Nació en Loja, Ecuador en 1982. Obtuvo el título de Ingeniero Eléctrico en la Escuela Politécnica Nacional, Ecuador en el 2007. En el año 2016, obtuvo el título de Doctor en Ingeniería Eléctrica en la Universidad Nacional de San Juan, Argentina. Actualmente trabaja en el Operador Nacional de Electricidad CENACE. Sus áreas de investigación son: Mercados de Energía, Confiabilidad, Análisis de la seguridad y vulnerabilidad.

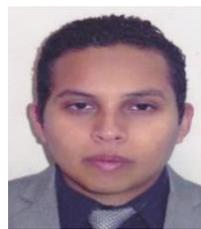

**WalterVargas**.- Nació en Guayaquil, Ecuador en 1984- Recibió sus títulos de Ingeniero en Electricidad, especialización Potencia (2007) en la Escuela Superior Politécnica del Litoral y el de Máster en Sistemas de Energía Eléctrica (2013)en la Universidad de Sevilla. Entre 2013 y el 2017 trabajó en la sección de Estudios Eléctricos del Departamento de Centro de Operaciones de CELEC EP –Transelectric. Actualmente se desempeña como profesor universitario a tiempo completo en la EPN. Sus áreas de interés incluyen la optimización, confiabilidad, evaluación de vulnerabilidad en tiempo real y el desarrollo de Smart Grids.

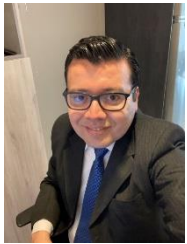

**Diego Echeverría**.- Recibió su título de Ingeniero Eléctrico de la Escuela Politécnica Nacional de Quito, en 2006. En el año 2021, obtuvo el título de Doctor en Ingeniería Eléctrica en la Universidad Nacional de San Juan, Argentina. Actualmente trabaja en el Operador Nacional de Electricidad CENACE de Ecuador como Gerente Nacional de Desarrollo Técnico. Sus áreas de interés son: Estabilidad de Sistemas de Potencia en Tiempo Real, Sistemas de medición sincrofasoriales PMU's y Control de Emergencia de Sistemas de Potencia.